\documentclass[letterpaper,twocolumn,english,aps,floatfix,showpacs,tightenlines,superscriptaddress,groupedaddress]{revtex4}
\usepackage[]{fontenc}
\usepackage[latin9]{inputenc}
\usepackage{amsmath}
\usepackage{graphicx}
\usepackage{amssymb}

\makeatletter

\newcommand{\lyxdot}{.}

\@ifundefined{textcolor}{}
{%
 \definecolor{BLACK}{gray}{0}
 \definecolor{WHITE}{gray}{1}
 \definecolor{RED}{rgb}{1,0,0}
 \definecolor{GREEN}{rgb}{0,1,0}
 \definecolor{BLUE}{rgb}{0,0,1}
 \definecolor{CYAN}{cmyk}{1,0,0,0}
 \definecolor{MAGENTA}{cmyk}{0,1,0,0}
 \definecolor{YELLOW}{cmyk}{0,0,1,0}
 }

\usepackage{hyperref}

\newcommand{\tr}{\mathrm{tr}}

\newcommand{\1}{\leavevmode{\rm 1\ifmmode\mkern  -4.8mu\else\kern -.3em\fi I}}

\makeatother

\usepackage{babel}

\begin{document}

\title{Universal equilibrium distribution after a small quantum quench}

\author{Lorenzo Campos Venuti}

\affiliation{Institute for Scientific Interchange (ISI), Viale S. Severo 65, I-10133
Torino, Italy }

\author{Paolo Zanardi}

\affiliation{Department of Physics and Astronomy and Center for Quantum Information
Science \& Technology, University of Southern California, Los Angeles,
California 90089-0484, USA}

\affiliation{Institute for Scientific Interchange (ISI), Viale S. Severo 65, I-10133
Torino, Italy }
\begin{abstract}
A sudden change of the Hamiltonian parameter drives a quantum system
out of equilibrium. For a finite-size system, expectations of observables
start fluctuating in time without converging to a precise limit. A
new equilibrium state emerges only in probabilistic sense, when the
probability distribution for the observables expectations over long
times concentrate around their mean value. In this paper we study
the full statistic of generic observables after a small quench. When
the quench is performed around a regular (i.e.~non-critical) point
of the phase diagram, generic observables are expected to be characterized
by Gaussian distribution functions (``good equilibration''). Instead,
when quenching around a critical point a new universal double-peaked
distribution function emerges for relevant perturbations. Our analytic
predictions are numerically checked for a non-integrable extension
of the quantum Ising model. 
\end{abstract}


\maketitle

\paragraph*{Introduction}

Imagine to prepare a closed quantum system in a given initial state
$\rho_{0}$ and let it evolve freely. After waiting a sufficiently
long time an equilibrium, average state $\overline{\rho}$ emerges.
Because of the unitary nature of the dynamics, in a finite system,
the evolved state $\rho\left(t\right)$ cannot converge to $\overline{\rho}$
either in the strong nor in the weak topology %
\footnote{For a different point of view see \cite{cramer08}.%
}. Equilibration in isolated quantum systems only emerges in a probabilistic
fashion. We say that the observable $O$ equilibrates to $\overline{O}$
if the expectation value $\langle O\left(t\right)\rangle$ spends
most of the times close to its average $\overline{O}$. In other words,
$\langle O\left(t\right)\rangle$ is seen as a random variable equipped
with the (uniform) measure $dt/T$ in the interval $t\in\left[0,T\right]$
where $T$ is the total observation time which will be sent to infinity.
The probability distribution of $O$ is $P\left(o\right):=\overline{\delta\left(o-\langle O\left(t\right)\rangle\right)}$,
where the bar refers to temporal averages: $\overline{f}:=\lim_{T\to\infty}T^{-1}\int_{0}^{T}f\left(t\right)dt$.
Broadly speaking concentration phenomena for $P\left(o\right)$ correspond
to quantum equilibration. The average value of a generic observable
is readily obtained as $\overline{O}:=\overline{\langle O\left(t\right)\rangle}=\tr\left(\overline{\rho}O\right)$,
an equation that defines the equilibrium state to be $\overline{\rho}=\overline{\rho\left(t\right)}$.
Equilibration however, is related to the concentration of the distribution
$P\left(o\right)$, a convenient definition of which is encoded in
the variance $\Delta O^{2}$. In Ref.~\cite{reimann08,winter2} it
has been shown that the variance of any observable is bounded by the
purity of the equilibrium state $\mathcal{P}\left(\overline{\rho}\right):=\tr\left(\overline{\rho}^{2}\right)$:
This is an encouraging result, if $\mathcal{P}\left(\overline{\rho}\right)$
is small one has equilibration for every observable. Equilibration
should depend on the dynamic and possibly on the initial state, not
on the specific observable.

A convenient setting to probe quantum equilibration is that of a sudden
quench. The system is initialized in the ground state of some Hamiltonian
$H_{1}$, and then evolved unitarily with a small perturbation $H_{2}=H_{1}+\delta\lambda V$.
This situation is compelling both from a theoretical and an experimental
point of view thanks to the recent advances in cold atoms technology
\cite{newtons_cradle,sadler06,weiler08}.

In this paper we will analyze the full statistic of a generic observable
$P\left(o\right)$ after a small quench. For small quenches performed
around a regular (i.e.~non-critical point) the expected distribution
$P\left(o\right)$ is Gaussian in the generic case. Equilibration
is achieved in a standard fashion. Instead for quenches performed
around a critical point the distribution of generic observables tend
to a new, universal double peaked function which we are able to compute.

This behavior has been first demonstrated in \cite{bat1} for a particular
observable (the Loschmidt echo) on the hand of an exactly solvable
model (Ising model in transverse field). Here we show that the scenario
first advocated in \cite{bat1} is in fact general to small quenches
for sufficiently relevant perturbations.

\paragraph*{Critical scaling of the time-averaged state }

Here we consider the equilibrium distribution for small quench. When
the quench is small one can either expand the eigenvectors of the
evolution Hamiltonian $H_{2}$ with perturbation $+\delta\lambda V$
or expand the initial state with respect to a perturbation $-\delta\lambda V$.
We take the latter point of view. Let the $t>0$ Hamiltonian be $H_{2}=\sum_{n}E_{n}|n\rangle\langle n|$.
The initial state $|\psi_{0}\rangle$ is the ground state of $H_{1}=H_{2}-\delta\lambda V$.
Then\[
|\psi_{0}\rangle=|0\rangle+\delta\lambda\sum_{n\neq0}\frac{\langle n|V|0\rangle}{E_{n}^{\left(2\right)}-E_{0}^{\left(2\right)}}|n\rangle+O\left(\delta\lambda^{2}\right)\]
 (note the \emph{plus} sign in $V$). If the spectrum is non-degenerate
the equilibrium state has the form $\overline{\rho}=\sum_{n}p_{n}|n\rangle\langle n|$
\cite{reimann08,winter2,bat1}. The weights, up to second order in
the quench potential, are given by\begin{align}
p_{0}=\left|\langle0|\psi_{0}\rangle\right|^{2} & =1-\delta\lambda^{2}\sum_{m\neq0}\frac{\left|\langle\psi_{m}|V|\psi_{0}\rangle\right|^{2}}{\left(E_{m}^{\left(2\right)}-E_{0}^{\left(2\right)}\right)^{2}}\nonumber \\
p_{n}=\left|\langle n|\psi_{0}\rangle\right|^{2} & =\delta\lambda^{2}\frac{\left|\langle0|V|n\rangle\right|^{2}}{\left(E_{0}^{\left(2\right)}-E_{n}^{\left(2\right)}\right)^{2}}\,,\,\, n\neq0.\label{eq:pn}\end{align}
 Note that up to the same order, the purity of the equilibrium state
is given by $\tr\left(\overline{\rho}^{2}\right)=p_{0}^{2}$. The
weight $p_{0}$ is precisely the square of the well studied ground
state fidelity $F=\left|\langle0|\psi_{0}\rangle\right|$ \cite{zapa06,zagioco07,zhou08,gurev08}
and its scaling properties are well known \cite{LCV07}. If the perturbing
potential is extensive and the quench is done around a regular (i.e.~non-critical)
point $F\sim\exp\left(-\mathrm{const}\times\delta\lambda^{2}L^{d}\right)$
where $d$ is the spatial system dimension. Instead for quenches at
a critical point $F\sim\exp\left(-\mathrm{const}\times\delta\lambda^{2}L^{2\left(d+\zeta-\Delta_{V}\right)}\right)$,
$\zeta$ is the dynamical critical exponent and $\Delta_{V}$ is the
scaling dimension of the perturbation $V$. Indeed it is intuitively
clear that by shrinking $\delta\lambda$ at will one should be able
to transfer most of the spectral weight to $p_{0}$, a limit in which
the purity is large. The above scalings tell us that we must have
$\delta\lambda\ll L^{-Q}$ with $Q=d/2$ ($Q=d+\zeta-\Delta_{V}=1/\nu$)
in the regular (critical) case. These are the regimes of small quench
characterized by a large purity and hence large variances for generic
observables. In other words poor equilibration.

However the distribution of the $p_{n}$ for critical and regular
quenches are radically different. As we will see, this has direct
consequences to the \emph{general form} of the distribution of generic
observables.

In case of a critical quench there exist modes with vanishing energy:
$E_{k}-E_{0}=vk^{\zeta}$ where $k$ now is a quasi-momentum label.
According to Eq.~(\ref{eq:pn}) the corresponding weight $p_{k}$
becomes large and might even (apparently) diverge when $k\to0$. In
a finite system with periodic boundary conditions the momenta are
quantized as $k=2\pi n/L$, then one would infer that, for a certain
weight $p_{1}\sim\delta\lambda^{2}L^{2\zeta}$. This, however is not
the correct scaling as we did not include the scaling of the matrix
element. To find the exact scaling we can reason as follows. Define
the functions $M\left(E_{n}\right):=\delta\lambda^{2}\left|\langle0|V|n\rangle\right|^{2}$,
and $p\left(E_{n}\right):=p_{n}$. With the help of the density of
states $\rho\left(E\right)=\tr\delta\left(E-H\right)$, one can write
the fidelity susceptibility $\chi$ as \begin{equation}
\chi=\delta\lambda^{2}\sum_{m\neq0}\frac{\left|\langle m|V|0\rangle\right|^{2}}{\left(E_{m}-E_{0}\right)^{2}}=\int_{E_{1}}^{E_{\mathrm{max}}}\frac{M\left(E\right)}{\left(E-E_{0}\right)^{2}}\rho\left(E\right)dE.\label{eq:chi}\end{equation}
 We are interested in the scaling properties of $M\left(E\right)$
after a rescaling of the energy. At criticality it is natural to assume
that $M\left(E\right)$ be an homogeneous function at the lower edge:
$M\left(E\right)\sim\left(E-E_{0}\right)^{\alpha}$. Instead the product
$\rho\left(E\right)dE$ is invariant under rescaling of the energy.
The scaling of the fidelity susceptibility is known \cite{LCV07}:
$\chi\sim L^{2\left(d+\zeta-\Delta_{V}\right)}\sim E^{-2\left(d+\zeta-\Delta_{V}\right)/\zeta}$
so, from $\chi\sim E^{\alpha-2}$, we obtain $\alpha=2\left(\Delta_{V}-d\right)/\zeta$.
Using the fact that, for the operator driving the transition $\Delta_{V}=d+\zeta-1/\nu$
\cite{schwandt09}, we obtain\begin{equation}
p\left(E\right)\sim\delta\lambda^{2}E^{-2/\left(\zeta\nu\right)}.\label{eq:pe_scaling}\end{equation}
 In this last equation the energy is measured from the ground state,
so that, being the system critical, $E$ can be arbitrarily close
to zero in the large size limit. The prediction Eq.~(\ref{eq:pe_scaling})
agrees with an explicit calculation on the quantum Ising model ($p\left(\omega\right)=2c\left(\omega\right)$
in \cite{bat1})

As a by-product of this analysis we obtain $\langle0|V|k\rangle\sim\delta\lambda L^{d-\Delta_{V}}=\delta\lambda L^{-\zeta+1/\nu}$.
Note that here $V$ is the extensive perturbation. If $V=\sum_{x}V\left(x\right)$,
for the intensive component we get\begin{equation}
\langle0|V\left(x\right)|k\rangle\sim\delta\lambda L^{-\Delta_{V}}=\delta\lambda L^{-\zeta-d+1/\nu}.\label{eq:matrix_scaling}\end{equation}
 Equation (\ref{eq:matrix_scaling}) is in agreement with the analysis
of \cite{barankov09} performed on the sine-Gordon model. In that
case $d=\zeta=1$ and one gets $\langle0|\cos\left(\beta\phi\left(x\right)\right)|k\rangle\sim L^{-2+1/\nu}$.
In fact formula (12) of \cite{barankov09} can be written as $\langle0|\cos\left(\beta\phi\left(x\right)\right)|k\rangle\sim L^{-K}$
where $K=2-1/\nu$ is the scaling dimension of the cosine term.

The content of equation (\ref{eq:pe_scaling}) is the following. For
a relevant perturbation ($d+\zeta>\Delta_{V}$) of a critical point
some spectral weights $p_{n}$ tend to be large. At finite size, the
lowest modes have energy, $E_{n}=v\left(2\pi n/L\right)^{\zeta}$
so that $p_{n}\sim\delta\lambda^{2}L^{2/\nu}$. In practice, since
in the region of validity of perturbation theory, $p_{0}$ is already
{}``large'', the sum rule $\sum_{n}p_{n}=1$ constrains to have
only very few $p_{n}$ appreciably different from zero. We expect
this scenario to be more pronounced for strongly relevant perturbations,
in other words when the exponent $2/\nu$ is large. When this is the
case, the sum rule can be saturated by taking a very small number
of terms $n_{\mathrm{max}}$: $1=\sum_{n}p_{n}\approx\sum_{n=0}^{n_{\mathrm{max}}-1}p_{n}$.
In our numerical simulations (see below) we have verified that for
a case with $\nu=1$ the sum rule is already saturated by taking as
little as three terms i.e.~$n_{\mathrm{max}}=3$. Moreover most of
the weight is splitted between $p_{0}$ and $p_{1}$, while $p_{2}$
is already orders of magnitude smaller.

The same considerations can clearly be drawn for the amplitudes $c_{n}=\langle n|\psi_{0}\rangle=\delta\lambda\langle n|V|0\rangle/\left(E_{0}-E_{n}\right)+O\left(\delta\lambda^{2}\right)$
for $n>0$, for which $p_{n}=\left|c_{n}\right|^{2}$. Defining the
function $c\left(E_{n}\right)=c_{n}$ with the same reasoning as above,
one sees that, for $E\to0$, $c\left(E\right)\sim\delta\lambda E^{-1/\left(\zeta\nu\right)}$.
Alternatively, for some low lying excitations with quasi-momentum
$k$, $c_{k}=\langle k|\psi_{0}\rangle\sim\delta\lambda L^{1/\nu}$.
Since $c\left(E\right)$ is a rapidly decreasing function, and because
of the sum rule for the $c_{n}$, one obtains a good approximation
for the time evolved wave-function by just resorting to very few,
$n_{\mathrm{max}}$, amplitudes: $|\psi\left(t\right)\rangle\approx\sum_{n=0}^{n_{\mathrm{max}}-1}c_{n}e^{-itE_{n}}|n\rangle$.

\paragraph*{Equilibrium distribution for small quenches}

\begin{figure}
\noindent \begin{centering}
\includegraphics[width=4.2cm]{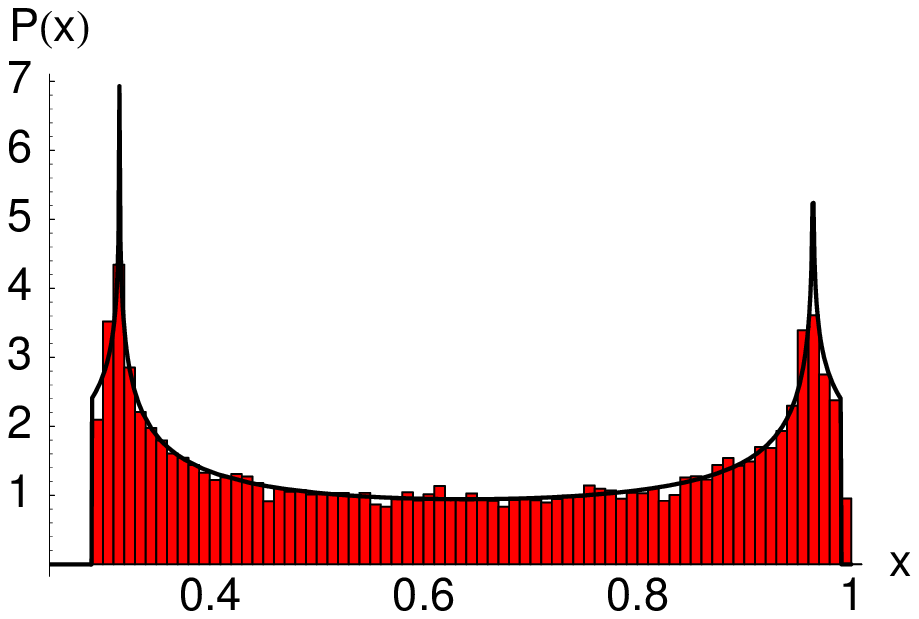} \includegraphics[width=4.2cm]{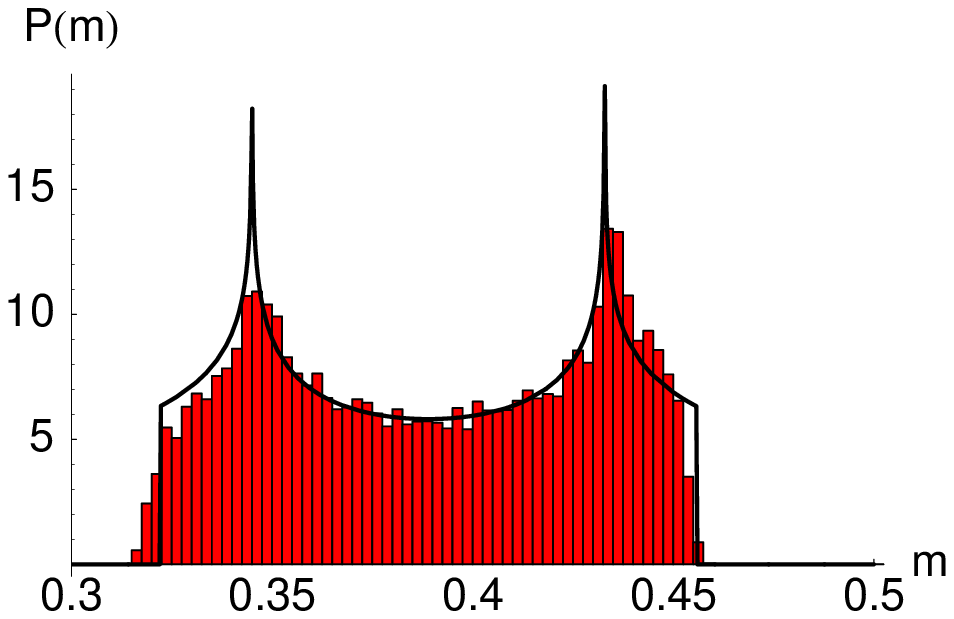} 
\par\end{centering}

\caption{(Color online) Probability distributions for a small quench around
a critical point. $P\left(x\right)=\overline{\delta\left(x-\mathcal{L}\left(t\right)\right)}$,
$P\left(m\right)=\overline{\delta\left(m-\langle\sigma_{1}^{z}\left(t\right)\rangle\right)}$
refer to the Loschmidt echo (upper panel) and magnetization respectively
(lower panel). The thick lines are obtained using the prediction of
Eq.~(\ref{eq:bat}) using only the three largest weights. Note the
large spread of the distributions compared with their total support:
$P\left(x\right)\in\left[0,1\right]$ and $P\left(m\right)\in\left[-1,1\right]$.
Parameters $L=16,\,\kappa_{1}=\kappa_{2}=0.4,\, h_{1}=0.218$, $\delta h=h_{2}-h_{1}=0.04$,
are close to criticality (see \cite{beccaria06}). The data are obtained
by Lanczos diagonalization of Eq.~(\ref{eq:TAM}) keeping as many
lowest energy vectors until the sum rule $\sum_{n=0}^{n_{\mathrm{max}}-1}p_{n}\simeq1$
was satisfied within prescribed accuracy. \label{fig:batman}}

\end{figure}

Let us now illustrate what are the consequences of these findings
on the equilibration. Consider the time evolution of a generic observable
$\langle O\left(t\right)\rangle$ . We will also give results for
the Loschmidt echo (LE) as it is attracting an increasing amount of
attention \cite{prosen98,pastawski01,quan06,rossini_base07,rossini07,PZ07,silva08}.
The Loschmidt echo is defined as $\mathcal{L}\left(t\right)=\left|\langle\psi_{0}|e^{-itH_{2}}|\psi_{0}\rangle\right|^{2}$.
Note that, as pointed out in \cite{bat1} the LE can be written as
the expectation value of a particular observable $\langle O_{\mathcal{L}}\left(t\right)\rangle$
with $O_{\mathcal{L}}$ given by $O_{\mathcal{L}}=|\psi_{0}\rangle\langle\psi_{0}|$.
Expanding \emph{$\mathcal{L}\left(t\right)$} and \emph{$\langle O\left(t\right)\rangle$}
in the eigenbasis of $H_{2}$ we obtain:

\begin{align}
\mathcal{L}\left(t\right) & =\overline{\mathcal{L}}+\sum_{m>n}2p_{n}p_{m}\cos\left[t\left(E_{n}-E_{m}\right)\right]\label{eq:LE_series}\\
O\left(t\right) & =\overline{O}+\sum_{n\neq m}\langle n|O|m\rangle c_{m}\overline{c_{n}}e^{-it\left(E_{m}-E_{n}\right)}\nonumber \\
 & =\overline{O}+\sum_{n>m}2\langle n|O|m\rangle c_{m}c_{n}\cos\left[t\left(E_{m}-E_{n}\right)\right].\label{eq:O_series}\end{align}
 Where in the last line we assumed that both the observables and the
wavefunctions are real as happens in most cases. As we have seen,
for a small quench around criticality both $c_{n}$ and $p_{n}$ will
be rapidly decreasing after their maximal value (in modulus), and
a good approximation to Eqns.~(\ref{eq:LE_series}) and (\ref{eq:O_series})
can be obtained by retaining only few terms. We have observed that
the following minimal prescription retaining only the three largest
components works fairly well:\begin{equation}
F\left(t\right)=\overline{F}+A\cos\left(\omega_{A}t\right)+B\cos\left(\omega_{B}t\right).\label{eq:bat}\end{equation}
 For instance $A=2p_{0}p_{1},\, B=2p_{0}p_{2}$ for the Loschmidt
echo while $A=2O_{0,1}c_{0}c_{1},\, B=2O_{0,2}c_{0}c_{2}$ for a more
generic observable $O$. The distribution function related to the
time-signal Eq.~(\ref{eq:bat}), $P\left(f\right)=\overline{\delta\left(f-F\left(t\right)\right)}$,
has been computed exactly in Ref.~\cite{bat1}. $P\left(f\right)$
is a symmetric function around the mean $\overline{F}$ supported
in $\left[\overline{F}-\left|\left|A\right|+\left|B\right|\right|,\overline{F}+\left|\left|A\right|+\left|B\right|\right|\right]$
with logarithmic divergences at $f=\overline{F}\pm\left|\left|A\right|-\left|B\right|\right|$
(see Fig.~\ref{fig:batman}).

This scenario can be summarized as follows: \emph{For small quench
around a critical point, generic observables equilibrate only very
poorly. The distribution function for a generic observable is a double
peaked distribution with a relatively large mean, a behavior completely
different from the Gaussian one.}

To complete the analysis let us now discuss the case of a small quench
in a regular point of the phase diagram. At regular points there are
no gapless excitations and the weights are bounded by $p\left(E\right)\le M\left(E\right)/\Delta^{2}$
where $\Delta$ is the smallest gap. Since the theory is not scale-invariant
$M\left(E\right)$ will not be an homogeneous function, and in particular
will not display any singularity. The picture then is the following:
In the perturbative regime ($\delta\lambda^{2}L^{d}\lesssim1$) we
still have a {}``large'' lowest weight, but beside $p_{0}$ no other
$p_{n}$ dominates and the sum rule $\sum_{n}p_{n}=1$ is saturated
only recurring to a relatively large bunch of $p_{n}$s.

In general predicting the precise behavior of observables in this
case will be difficult as one needs to have knowledge of many different
weights in Eqns.~(\ref{eq:LE_series}) and (\ref{eq:O_series}).
However we can give a simple argument to expect a Gaussian behavior
for \emph{generic} case. As we have argued, the sum in Eq.~(\ref{eq:O_series})
contains now many terms. If the energy differences $E_{n}-E_{m}$
are rationally independent, along the time evolution, each variable
$X_{n,m}:=2\langle n|O|m\rangle c_{m}c_{n}$ will span uniformly the
interval $\left[-2\left|\langle n|O|m\rangle c_{m}c_{n}\right|,2\left|\langle n|O|m\rangle c_{m}c_{n}\right|\right]$.
As long as the variables $X_{n,m}$ can be considered \emph{independent},
$O\left(t\right)-\overline{O}$ can be thought of as a sum of independent
random variables. Since, as we have seen, the sum is made over many
variables, the central limit theorem applies and the resulting distribution
will be Gaussian. This argument can fail when the variables cannot
be taken as independent. This can happen, for instance, when a certain
observable is pushed toward its maximum or minimal value by the action
of some field. Consider for example the case of a transverse magnetization
$\sigma_{j}^{z}$ in presence of a high field $-h\sum_{i}\sigma_{i}^{z}$.
For increasing $h$ the mean of $\langle\sigma_{j}^{z}\left(t\right)\rangle$
will be pushed towards one. Since $\langle\sigma_{j}^{z}\left(t\right)\rangle$
is supported in $\left[-1,1\right]$ the corresponding distribution
can cease to be Gaussian as its mean is pushed against the (upper)
border of its support. In this case the distribution function will
look like a {}``squeezed'' Gaussian. A similar effect has been observed
to take place to the Loschmidt echo in Ref.~\cite{bat1} when the
system size becomes the largest scale of the system. In any case,
however, if the variables cannot be considered as independent, any
possible distribution function (and not only a squeezed Gaussian)
can arise.

\begin{figure}
\noindent \begin{centering}
\includegraphics[width=4.2cm]{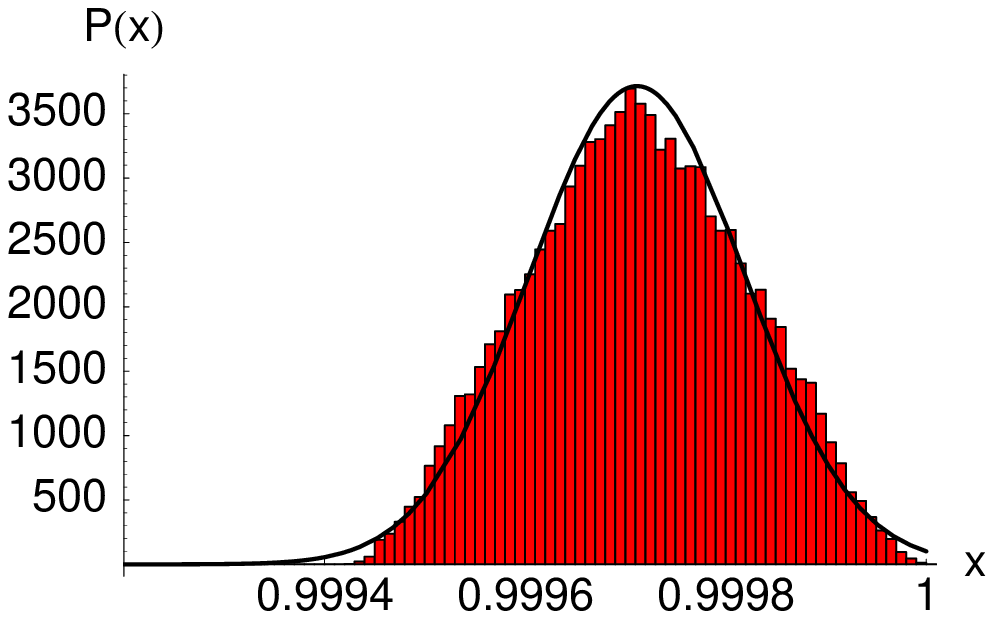}
\includegraphics[width=4.2cm]{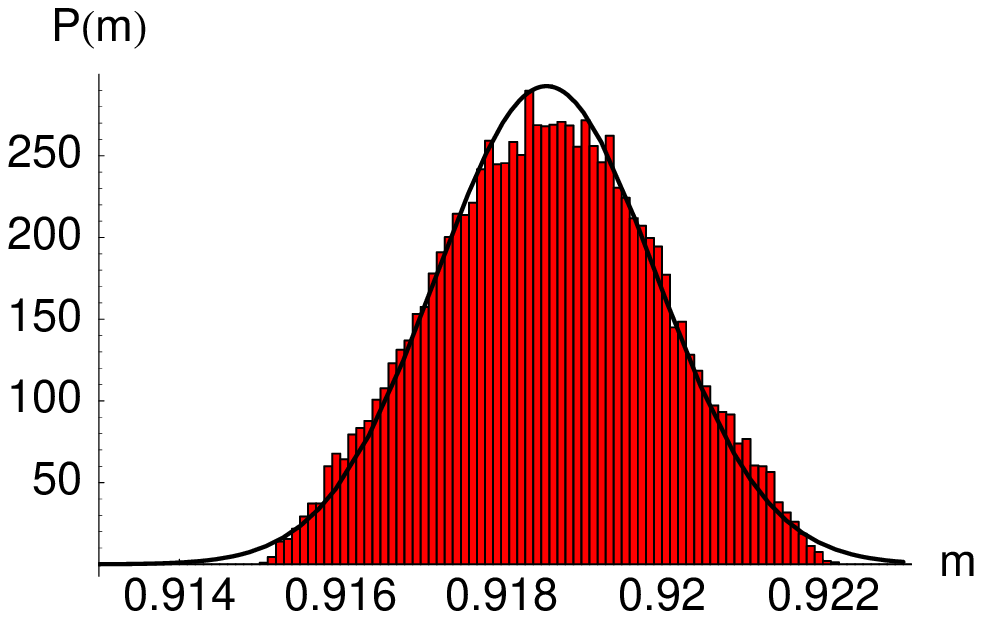} 
\par\end{centering}

\caption{(Color online) Probability distributions for a small quench around
a regular point. $P\left(x\right)=\overline{\delta\left(x-\mathcal{L}\left(t\right)\right)}$,
$P\left(m\right)=\overline{\delta\left(m-\langle\sigma_{1}^{z}\left(t\right)\rangle\right)}$
refer to the Loschmidt echo (upper panel) and magnetization respectively
(lower panel). The thick lines are Gaussian with the same mean and
standard deviation. The quench is performed in the paramagnetic phase,
parameters are $L=12,\,\kappa_{1}=\kappa_{2}=0.3,\, h_{1}=1.4$, $\delta h=0.04$.
Note the very small spread of the distributions. The data are obtained
by full diagonalization of the Hamiltonian Eq.~(\ref{eq:TAM}). Histograms
are obtained by sampling 40000 random times in an interval $t\in\left[0,T\right]$
with $T=16000$. \label{fig:Gaussian}}

\end{figure}

\paragraph*{Numerical test}

We will now check our predictions on the hand of a non-integrable
model. As a test model we chose to use the so called TAM Hamiltonian
(transverse axial next-nearest-neighbor Ising model). The Hamiltonian
is\begin{equation}
H=-\sum_{i=1}^{L}\left(\sigma_{i}^{x}\sigma_{i+1}^{x}-\kappa\sigma_{i}^{x}\sigma_{i+2}^{x}+h\sigma_{i}^{z}\right),\label{eq:TAM}\end{equation}
 and periodic boundary conditions are used ($\sigma_{L+i}^{x}=\sigma_{i}^{x}$).
A positive $\kappa$ frustrates the order in the $\sigma^{x}$ direction.
The reason for our choice is, at least, twofold: i) The TAM is a non-integrable
generalization of the one-dimensional quantum Ising model for which
results are already available \cite{bat1}. ii) The model Eq.~(\ref{eq:TAM})
has only a discrete $\mathbb{Z}_{2}$ symmetry ($P_{z}=\prod_{i}\sigma_{i}^{z}$),
consequently the ground state lives in a large $d_{GS}=2^{L-1}$ dimensional
space. In practice $d_{GS}$ is the effective Hilbert space dimension,
and we would like it to be as large as possible. For instance, after
a quench the purity of the equilibrium state is bounded by $\tr\left(\overline{\rho}^{2}\right)\ge d_{GS}^{-1}$.
This is to be contrasted with other models used in the literature
with larger symmetry groups (i.e.~$SU\left(2\right)$) for which
the dimension of the block containing the ground state is still exponential
in $L$ but considerably reduced with respect to to that of the full
Hilbert space $2^{L}$.

The model Eq.~(\ref{eq:TAM}) displays 4 phases (see for instance
\cite{peschel81,allen01,beccaria06,beccaria07} and references therein),
ferromagnetic $++++$, antiphase $++--$, paramagnetic, and a floating
phase with algebraically decaying spin correlations. In particular,
for small frustration $\kappa\le1/2$, increasing the external field
$h$ there is a transition from ferromagnetic to paramagnetic. This
transition is believed to fall in the Ising universality class, and
so the critical theory is described by a conformal field theory with
central charge $c=1/2$ and $d=\zeta=\nu=1$. We performed our numerical
simulation for the critical quench on this critical line.

We will illustrate our findings for two particular yet physically
well motivated observables; the Loschmidt echo $\mathcal{L}\left(t\right)=\left|\langle\psi_{0}|e^{-itH_{2}}|\psi_{0}\rangle\right|^{2}$
and the transverse magnetization $m\left(t\right)=\langle\psi_{0}\left(t\right)|\sigma_{i}^{z}|\psi_{0}\left(t\right)\rangle$.

Since $d=\zeta=\nu=1$, according to Eq.~(\ref{eq:pe_scaling}),
we expect a strong divergence at low energy: $p\left(E\right)\sim E^{-2}$.
Consequently we expect very few $p_{n},\, n>0$ to have non-negligible
weight, and so Eq.~(\ref{eq:bat}) to be a valid approximation. Indeed
the results based on numerical diagonalization compare well with the
prediction based on Eq.~(\ref{eq:bat}) (Fig.~\ref{fig:batman}).
Note the very large spread of the distributions compared to their
total support: {}``poor equilibration''. 

For comparison we performed similar numerical simulation for a small
quench in a regular point of the phase diagram. As expected the resulting
distribution functions are approximately Gaussian (Fig.~\ref{fig:Gaussian}).
Note the very small variances of the distributions already for a relatively
short size: {}``good equilibration''

\paragraph*{Conclusions}

In this paper we investigated the detailed structure of equilibration
after a small quench, i.e.~the system is initialized in the ground
state of a given Hamiltonian $H_{1}$ and then let evolve with a slightly
perturbed Hamiltonian $H_{2}=H_{1}+\delta\lambda V$. In the limit
$\delta\lambda\to0$ equilibration is trivial in that for all observables
$P\left(o\right)=\overline{\delta\left(o-\langle O\left(t\right)\rangle\right)}^{t}=\delta\left(o-\langle O\rangle\right)$.
However this limit is approached very differently depending on whether
Hamiltonian $H_{1}$ is critical or not. For quenches around a regular
point of the phase diagram the expected distribution for generic observables
is a Gaussian one. Equilibration arises in the most standard fashion.
Instead for small quenches around a critical point the situation is
radically different. The distribution function for generic observables
$P\left(o\right)$ tends to universal double-peaked function for relevant
perturbations.

The key step to obtain these results is to characterize the overlaps
$c_{n}=\langle n|\psi_{0}\rangle$ between the initial state $|\psi_{0}\rangle$
and quenched Hamiltonian eigenstates $|n\rangle$. We have shown that,
at criticality, the function $c\left(E_{n}\right)=c_{n}$ ($E_{n}$
eigenenergy) decays very rapidly: $c\left(E\right)\sim E^{-1/\left(\zeta\nu\right)}$
and this in turns generically implies the observed double-peaked distributions.
The analytical predictions have been checked numerically on the hand
of a non-integrable extension of the quantum Ising model.

LCV acknowledges support from European project COQUIT under FET-Open
grant number 2333747 and PZ from NSF grants PHY-803304, DMR-0804914.
\bibliographystyle{apsrev}
\bibliography{ref_le}

\end{document}